\documentstyle[12pt,epsf]{article}
\input{epsf}
\begin{document}
\begin{center}
{\Large\bf Parity shift and beat staggering structure of octupole
bands in a collective model for quadrupole--octupole deformed nuclei}\\
\bigskip\bigskip

N. Minkov$^{a,b,}$\footnote{E-mail: nminkov@inrne.bas.bg},
P. Yotov$^{a,b,}$\footnote{E-mail: pyotov@inrne.bas.bg},
S. Drenska$^{a,}$\footnote{E-mail: sdren@inrne.bas.bg},
and
W. Scheid$^{b,}$\footnote{E-mail:
Werner.Scheid@theo.physik.uni-giessen.de} \\
\bigskip

$^{a}${\em Institute of Nuclear Research and Nuclear Energy,
72 Tzarigrad Road, Sofia 1784, Bulgaria} \\
\smallskip
$^b${\em Institut f\"{u}r Theoretische Physik der
Justus-Liebig-Universit\"at, Heinrich-Buff-Ring 16, D--35392
Giessen, Germany}
\end{center}
\bigskip

\begin{abstract}
We propose a collective model formalism which describes
the strong parity shift observed in low-lying spectra
of nuclei with octupole deformations together with the fine
rotational band structure developed at higher angular momenta. The
parity effect is obtained by the Schr\"{o}dinger equation for
oscillations of the reflection asymmetric (octupole) shape between
two opposite orientations in an angular momentum dependent
double-well potential. The rotational structure is obtained by a
collective quadrupole-octupole rotation Hamiltonian. The
model scheme reproduces the complicated beat staggering patterns
observed in the octupole bands of light actinide nuclei. It
explains the angular momentum evolution of octupole spectra as the
interplay between the octupole shape oscillation (parity shift)
mode and the stable quadrupole-octupole rotation mode.
\end{abstract}

\medskip

\medskip

Short title: Parity shift and beat staggering structure ...

\newpage

\section{Introduction}
The appearance of reflection asymmetric shapes in atomic nuclei is
associated in the geometric model framework with the manifestation
of octupole degrees of freedom \cite{BM75}. The specific physical
characteristics of systems with reflection asymmetry are related
to the violation of the $\mathcal{R}$- symmetry ($\mathcal{R}$ is
the operator of a rotation by $\pi$ about an axis rectangular to a
symmetry axis of the system) and the $\mathcal{P}$- symmetry
[$\mathcal{P}$ is the space inversion (parity) operator]. It is
known that while these symmetries are violated separately, the
system can be still invariant with respect to the product operator
$\mathcal{P}\mathcal{R}$$^{-1}$ \cite{BM75}. Then the spectrum of
the system is characterized with the presence of energy bands in
which the parity changes alternatively with the angular momentum.

The best examples of such alternating parity bands, called also
octupole bands, are known in the region of light actinide nuclei
Rn, Ra and Th \cite{Cocks97,Th224,Th226}. Although the general
structure of the observed spectra unambiguously indicates the
presence of reflection asymmetry, the detailed analysis of
experimental data suggests the manifestation of a more complicated
collective dynamics. Thus, a strong parity shift effect is
observed in the region of low angular momenta $I<7-8$. The
negative parity levels appear above the neighboring even levels
with energy $E_{1^{-}}>E_{2^{+}}$, $E_{3^{-}}>E_{4^{+}}$, and so
on. Near $I\sim 8-10$ the shift effect rapidly decreases and
further at higher angular momenta (up to $I\sim 28$) a single
collective octupole band  with normally ordered  levels is formed.
In such a way the data suggest a specific evolution of
collectivity with angular momentum.

The properties of nuclei with octupole degrees of freedom have
been extensively studied within various geometric, algebraic and
microscopic model approaches in nuclear structure (for review see
\cite{BN96} and references therein). In particular, the
alternating parity bands have been described in several recent
works, most of them covering the low and medium angular momenta
$I<18-20$ \cite{DZ95,Zam01,clust} and some of them reaching the
higher region of $I\sim 28$ \cite{Ra03}. These works
provide specific model explanations of the properties of octupole
deformed nuclei. However, the dynamical mechanism which governs
the complete angular momentum evolution of octupole collectivity
remains to be understood. Still, the considerable change in the
structure of the spectrum from the low-spin region ($I<10-12$) to
the medium ($I\sim 12-16$) and higher ($I\sim 20-28$) angular
momenta needs an explanation. This is quite clear in
view of the ``beat'' odd-even ($\Delta I =1$) staggering patterns
observed in the octupole bands of light actinide nuclei
\cite{DBoct00}. They correspond to a zigzagging behavior of the
odd-even energy difference as a discrete function of $I$ with
changing amplitude and presence of one or more zero-amplitude
points. Another important circumstance that should be taken into
account is that the octupole deformation appears as a superposition
on the top of the quadrupole deformation \cite{BN96}. Therefore,
one should consider a system with a complicated quadrupole-octupole
shape. A prescription for the most general Hamiltonian of the
quadrupole-octupole deformed system is given in \cite {Ro82}.
It is based on the Pauli quantization procedure applied to a
generalized Bohr Hamiltonian, including the octupole degrees
of freedom, with a subsequent transformation to the body fixed
frame.

Having in mind the above remarks, we suggest that the {\em
complete structure}  of alternating parity bands can be
interpreted within the following general framework. In the
low-energy region the system can be characterized by oscillations
of the octupole shape between two opposite orientations, which we
call {\em soft octupole mode}, and by simultaneous rotations of
the entire quadrupole--octupole shape. The parity shift is then
the result of the tunnelling between the two reflection asymmetric
shape orientations separated by an angular momentum dependent
potential barrier \cite{Jolos93}. With the increase of the angular
momentum the energy barrier increases and suppresses respectively
the tunnelling effect and the shape oscillations. In such a way a
{\em stable} quadrupole--octupole shape is formed. At the high
angular momenta, the parity effect is completely reduced, and
there the properties of collective motion can be associated with
the rotation of a stable quadrupole--octupole shape
\cite{octahed01}.

The purpose of the present work is to provide a model formalism
capable to describe the strong parity shift in the low-lying
spectra together with the fine rotational band structure developed
at higher angular momenta. This is achieved by introducing a
double-well potential with an angular momentum dependent energy
barrier induced by a centrifugal term and by using the Hamiltonian
of the point-symmetry based Quadrupole-Octupole Rotation Model
(QORM) \cite{octahed01}.

In section 2, the formalism of the soft octupole mode is developed
and in section 3 the formalism of QORM is discussed. In section 4,
results of the application of both schemes, soft octupole mode and
QORM, to spectra of several light actinide nuclei are presented.
In section 5, concluding remarks are given.

\section{Soft octupole deformation and parity shift effect}

The model frame for the description of the parity shift effect in
nuclei with octupole deformations can be defined through the
following main assumptions:

i) The parity shift in octupole bands is determined by the
oscillation of the system with respect to a deformation variable
$\beta$ in a symmetric double-well potential centered at
$\beta=0$. We assume that $\beta$ is related to the axial octupole
deformation ($Y_{30}$).

ii) The potential barrier increases with the increase of the
angular momentum $I$ as the result of a centrifugal interaction
with a deformation dependent moment of inertia.

iii) The octupole deformation $\beta_{\mbox{\scriptsize min}}$ in
the minimum of the potential does not change with the angular
momentum. Therefore, with increasing $I$ the octupole shape
``stabilizes'' to a fixed octupole deformation.

iv) The parity shift can be extracted from the potential energy by
excluding the contribution of the rotation mode. The rotation
degrees of freedom can be considered separately without double
counting in the potential energy.

v) For given $I$ the intrinsic state is determined by the lowest
level in the potential whose parity $\pi$ together with the
$\mathcal{R}$- symmetry of the rotation wave function, $(-1)^I$,
conserves the total $\mathcal{P}\mathcal{R}$$^{-1}$-symmetry,
$s=\pi (-1)^I$. Thus for $I=$even the intrinsic state of the
system is determined by the first level $E_I^{(+)}$ in the
potential with $\pi =(+)$. For $I=$odd it is determined by the
second level $E_I^{(-)}$ with $\pi =(-)$.

The Hamiltonian for oscillations in the octupole shape (the soft
octupole mode) can be taken in the general form
\begin{equation}
H^{\mbox{\scriptsize oct}}_{\mbox{\scriptsize osc}}
=-\frac{\hbar^{2}}{2B_3}\frac{d^{2}}{{d \beta}^{2}}+ U_I(\beta)\ ,
\label{Schro}
\end{equation}
where $B_3$ is an effective octupole mass parameter. The potential
$U_I(\beta)$ is determined as follows. We consider the angular
momentum dependent potential
\begin{equation}
U(\beta , I)=\frac{1}{2}C{\beta}^{2}+
\frac{1}{2{\mathcal{L}}(\beta)}I(I+1) \ .
\label{UbetaI}
\end{equation}
The quantity ${\mathcal{L}}(\beta)$ can be associated to the
moment of inertia of the system taken in the form
${\mathcal{L}}(\beta)=d_1+d_2\beta^2$, where $d_1,\, d_2>0$ are
constants. Such a dependence can be explained through the moment
of inertia of an axial symmetric quadrupole-octupole shape,
${\mathcal{L}}^{(\mbox{\scriptsize quad+oct})}\sim
3B_2\beta^2_{\mbox{\scriptsize
quad}}+6B_3\beta^2_{\mbox{\scriptsize oct}}$, where
$\beta_{\mbox{\scriptsize quad}}$ and $\beta_{\mbox{\scriptsize
oct}}=\beta$ are the quadrupole and octupole deformation
parameters, respectively \cite{JPDav68}. For a stable quadrupole
deformation, $\beta_{\mbox{\scriptsize quad}}=\mbox{const}$,
${\mathcal{L}}^{(\mbox{\scriptsize quad+oct})}$ reduces to the
above form of ${\mathcal{L}}(\beta)$, and the potential
(\ref{UbetaI}) reads
\begin{equation}
U(\beta ,I)=\frac{1}{2}C{\beta}^{2}+\frac{X(I)}{d_1+d_2\beta^2} \
, \label{Ubeta}
\end{equation}
with $X(I)=I(I+1)/2$. For $X(I)/C>d_1^2/2d_2$, the shape of the
potential (\ref{Ubeta}) represents a double well with a potential
barrier centered at $\beta=0$. The term quadratic in $\beta$
generates oscillations of the reflection asymmetric (octupole)
shape between two opposite orientations of the system, providing a
parity shift effect in the collective energy. The second term in
(\ref{Ubeta}) corresponds to a centrifugal interaction generating
a potential barrier. The height of the barrier increases with
increasing $I$ and suppresses respectively the parity shift.

\begin{figure}[t]
\epsfxsize=9.cm   %or \epsfysize= HEIGHT cm
\centerline{\epsfbox{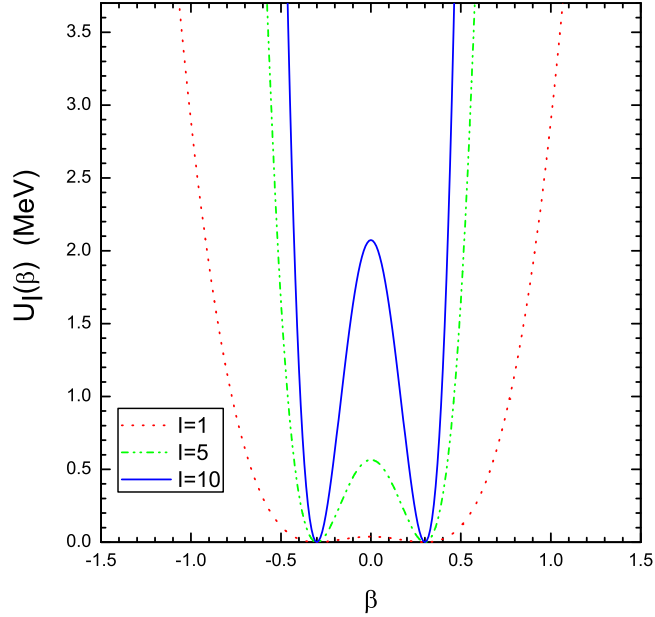}} \caption{The octupole deformation
potential (\protect\ref{octfix}) is plotted for three different
values of the angular momentum $I=1,5,10$. The parameters values
used in equation~(\protect\ref{octfix}) are given in figure~2.}
\label{fig:01}
\end{figure}

\begin{figure}[t]
\epsfxsize=9cm   %or \epsfysize= HEIGHT cm
\centerline{\epsfbox{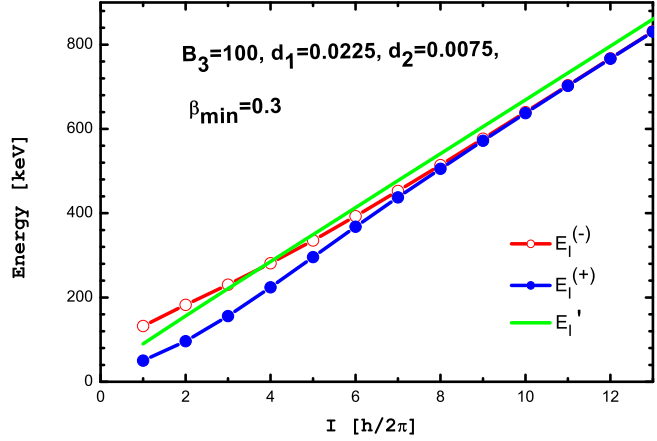}} \caption{The first and the
second energy levels $E_I^{(+)}$ and $E_I^{(-)}$ in the
potential (\protect\ref{octfix}) as well as the approximating
oscillator energy $E'_I$, equation~(\protect\ref{Eprim}), are shown as
functions of angular momentum. The parameters $B_3$, $d_1$ and
$d_2$ are given in $\hbar^2$/MeV, MeV$^{-1}$ and MeV$^{-1}$,
respectively, while $\beta_{\mbox{\scriptsize min}}$ is
dimensionless.} \label{fig:02}
\end{figure}

We remark that double-well potentials without angular momentum
dependence of the potential barrier have been considered in
earlier works \cite{KW1969,Leander}. Some general properties of
angular momentum dependent double-well potentials have been
discussed in references \cite{Jolos93,Jolos94}. A spin-dependent
potential with a centrifugal term similar to that in (\ref{Ubeta})
has been applied in \cite{Bizz0304,Bizz05} in reference to the
phase transitions in nuclei with octupole degrees of freedom.

Following the assumption iii), we consider that the potential
$U(\beta ,I)$ has a minimum at some fixed value of $\beta
=\beta_{\mbox{\scriptsize min}}$. The extremum condition $\left.
\frac{\partial}{\partial \beta}U(\beta
,I)\right|_{\beta_{\mbox{\scriptsize min}}}=0$ gives
\begin{equation}
C\equiv C(I)=\frac{2X(I)d_2}{(d_1+d_2\beta_{min}^2)^2}.
\label{CL}
\end{equation}
Thus, the requirement for a fixed potential minimum in the
octupole deformation imposes an angular momentum dependence of
the parameter $C$. The substitution of equation~(\ref{CL}) into
(\ref{Ubeta}), leads to the following form of the octupole
potential
\begin{eqnarray}
U(\beta,I) =\frac{X(I)}{d_1+d_2\beta^2}\left[1+
\frac{d_2(d_1+d_2\beta^2)}{(d_1+d_2\beta_{\mbox{\scriptsize
min}}^2)^2}\beta^2\right]\ . \label{octfixL}
\end{eqnarray}
To eliminate (according to assumption iv) the angular momentum
(rotation) dependence of the origin of the energy scale we
determine the potential as $U_{I}(\beta )
=U(\beta,I)-U(\beta_{\mbox{\scriptsize min}},I)$,
\begin{eqnarray}
U_{I}(\beta )=\frac{X(I)d_2^2(\beta^2-\beta_{\mbox{\scriptsize
min}}^2)^2} {(d_1+d_2\beta_{\mbox{\scriptsize min}}^2)^2
(d_1+d_2\beta^2)} . \label{octfix}
\end{eqnarray}
The evolution of the potential (\ref{octfix}) with the angular
momentum is demonstrated in figure~1. It is seen that, the increase
in the potential barrier with $I$ is associated with a respective
decrease in the width of the potential well. This corresponds to
an increasing stiffness of the system, and therefore, to a rapid
stabilization of the shape in the higher angular momenta.

\begin{figure}[t]
\epsfxsize=20cm   %or \epsfysize= HEIGHT cm
\centerline{\epsfbox{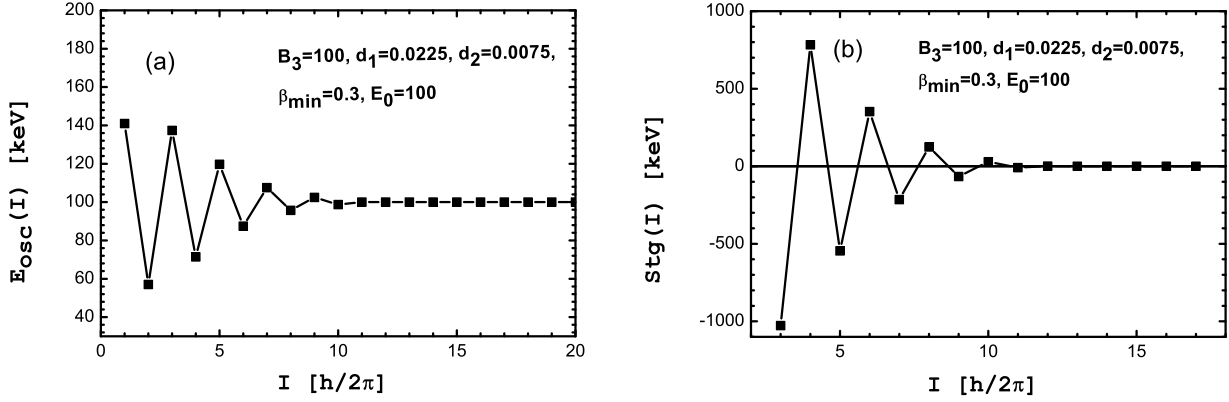}} \caption{Numerical behavior of the
octupole oscillation term $E_{\mbox{\scriptsize osc}}(I)$
(\protect\ref{Elowvib}) [part (a)] and its contribution to the
staggering quantity (\protect\ref{stag}) [part (b)]. $E_0$ is
given in keV, while the other parameters are as in figure 2.}
\label{fig:03}
\end{figure}

We determine the energy levels $E_I^{(+)}$ and $E_I^{(-)}$
[according to the assumption v)] in the potential $U_{I}(\beta )$
[equation~(\ref{octfix})] by solving numerically the Schr\"{o}dinger
equation with the Hamiltonian (\ref{Schro}). These energies still
carry a rotation input due to the narrowing of the potential
minima. As shown in figure 2, the both levels increase with $I$. To
estimate analytically this dependence we approximate the shapes of
the potential minima by $U'_{I}(\beta )= (1/2)C'(I)(|\beta |
-\beta_{\mbox{\scriptsize min}})^{2}$, with $C'(I)=\left.
\frac{\partial^{2}}{\partial\beta^{2}} U_{I}(\beta
)\right|_{\beta_{\mbox{\scriptsize min}}}$. The lowest energy
level in this potential is
\begin{equation}
E'_I=\frac{\hbar}{2}\sqrt{\frac{C'(I)}{B_3}}=
\frac{\hbar}{\sqrt{2B_3}}\frac{d_{2}\beta_{\mbox{\scriptsize
min}}} {(d_1+d_2\beta_{\mbox{\scriptsize min}}^2)^{3/2}}
\sqrt{I(I+1)}. \label{Eprim}
\end{equation}
It approximates the overall increase of the levels $E_I^{(+)}$ and
$E_I^{(-)}$ as demonstrated in figure 2. Thus, to {\em keep only the
parity dependence in the intrinsic energy} according to assumption
iv), and {\em not the part nearly linearly increasing with $I$}
one should set the energy of the octupole shape oscillations in
the form
\begin{equation}
E_{\mbox{\scriptsize osc}}(I)=E_0-\frac{1}{2}(-1)^I\delta E(I),
\label{Elowvib}
\end{equation}
where $\delta E(I)=E_I^{(-)}-E_I^{(+)}$ and $E_0$ is a constant.
Thus the intrinsic energy related to octupole degrees of freedom
is separated in a part independent of the angular momentum, and an
``alternating'' part, which takes into account the shift down (for
$I=$even) and the shift up (for $I=$odd) of the intrinsic
``octupole'' energy with respect to  $E_0$. So,
equation~(\ref{Elowvib}) approximates the {\em non-rotational} part of
the energy in the alternating parity bands.

The angular momentum dependence of $E_{\mbox{\scriptsize osc}}(I)$
is demonstrated in figure~3(a). We see that in the low spin region
it exhibits a strongly oscillating behavior. With increasing $I$
the oscillations rapidly vanish and $E_{\mbox{\scriptsize
osc}}(I)$ is reduced to the constant $E_0$. In figure~3(b) we
illustrate the contribution of $E_{\mbox{\scriptsize osc}}(I)$ to
the higher order staggering quantity
\begin{equation}
\mbox{Stg}(I)= 6\Delta E(I)-4\Delta E(I-1)-4\Delta E(I+1)+ \Delta
E(I+2)+\Delta E(I-2)\ , \label{stag}
\end{equation}
where $\Delta E(I)=E(I+1)-E(I)$. As it is seen from figure~3(b), for
$E(I)\equiv E_{\mbox{\scriptsize osc}}(I)$, this quantity shows a
well developed staggering pattern with a rapidly decreasing
magnitude similar to what is observed in the low spin region of
alternating parity bands in light actinide nuclei \cite{DBoct00}.
However, while the solutions of the Schr\"{o}dinger equation for
the Hamiltonian (\ref{Schro}) suggest a complete disappearance of
the parity shift and the respective staggering after some point,
the experimental data show \cite{DBoct00} more complicated
``beat'' patterns with the presence of further staggering regions
at higher angular momenta. This additional staggering regions can
not be explained by the octupole oscillation mode. Their presence
can be attributed to the properties of a rotating system with a
stable quadrupole-octupole deformation which will be considered in
the next section.

\section{The Quadrupole--Octupole Rotation Model}

The Quadrupole--Octupole Rotation Model (QORM) has been proposed
for the description of rotation motion in nuclear systems in the
presence of stable quadrupole--octupole shapes \cite{octahed01}.
The general model Hamiltonian is given in the form
\begin{equation}
\hat{H}_{\mbox{\scriptsize QORM}}=\hat{H}_{\mbox{\scriptsize
quad}}+\hat{H}_{\mbox{\scriptsize oct}}+
\hat{H}_{\mbox{\scriptsize qoc}} \ . \label{Hgen}
\end{equation}

The quadrupole rotation term
\begin{equation}
\hat{H}_{\mbox{\scriptsize quad}}= A\hat{I}^{2}+A'\hat{I}_{z}^{2}
\label{Hrot}
\end{equation}
provides the general energy scale for rotation motion of the
nucleus. The octupole Hamiltonian part
\begin{equation}
\hat{H}_{\mbox{\scriptsize oct}}=\hat{H}_{A_{2}}+
\sum_{r=1}^{2}\sum_{i=1}^{3}\hat{H}_{F_{r}(i)} \label{Hoct}
\end{equation}
is constructed by using the irreducible representations $A_{2}$,
$F_{1}(i)$ and $F_{2}(i)$ ($i=1$, 2, 3) of the octahedron $(O)$
point--symmetry group, where
\begin{equation}
\hat{H}_{A_{2}}={a}_{2}\frac{1}{4}
[(\hat{I}_x\hat{I}_y+\hat{I}_y\hat{I}_x)\hat{I}_z+
\hat{I}_z(\hat{I}_x\hat{I}_y+\hat{I}_y\hat{I}_x)] \ , \label{HA}
\end{equation}
\begin{eqnarray}
\hat{H}_{F_{1}(1)}=\frac{1}{2}{f}_{11}
(5\hat{I}_z^{3}-3\hat{I}_z\hat{I}^{2}) &\qquad &
\hat{H}_{F_{2}(1)}={f}_{21}\frac{1}{2}
[\hat{I}_z(\hat{I}_x^{2}-\hat{I}_y^{2})+
(\hat{I}_x^{2}-\hat{I}_y^{2})\hat{I}_z]
\nonumber \\
\hat{H}_{F_{1}(2)}=\frac{1}{2}{f}_{12}
(5\hat{I}_x^{3}-3\hat{I}_x\hat{I}^{2}) &\qquad &
\hat{H}_{F_2(2)}={f}_{22} (\hat{I}_x\hat{I}^{2}-\hat{I}_x^{3}-
\hat{I}_x\hat{I}_z^{2}-\hat{I}_z^{2}\hat{I}_x)
\label{HF1F2} \\
\hat{H}_{F_{1}(3)}=\frac{1}{2}{f}_{13}
(5\hat{I}_y^{3}-3\hat{I}_y\hat{I}^{2}) &\qquad &
\hat{H}_{F_2(3)}={f}_{23}
(\hat{I}_y\hat{I}_z^{2}+\hat{I}_z^{2}\hat{I}_y+
\hat{I}_y^{3}-\hat{I}_y\hat{I}^{2})\ . \nonumber
\end{eqnarray}
The different terms in $\hat{H}_{\mbox{\scriptsize oct}}$ (cubic
combinations of angular momentum operators in the body fixed
frame) represent the contribution of the various octupole shapes
to the rotation energy with a magnitude determined by the
parameters ${a}_{2}$ and ${f}_{r\, i}$ ($r=1,2;\, i=1,2,3$). The
last term in equation~(\ref{Hgen}) represents a higher order
quadrupole--octupole interaction \cite{octahed01}
\begin{equation}
\hat{H}_{\mbox{\scriptsize qoc}} =f_{\mbox{\scriptsize qoc}}
\frac{1}{I^{2}}(15\hat{I}_{z}^{5}-14\hat{I}_{z}^{3}\hat{I}^{2}+
3\hat{I}_{z}\hat{I}^{4}). \label{Hqoc}
\end{equation}

The energy corresponding to the diagonal part of
$\hat{H}_{\mbox{\scriptsize QORM}}$ in the states $|IK\rangle$
with collective angular momentum $I$ and intrinsic projection $K$
has the form
\begin{eqnarray}
E_{K}(I)&=&AI(I+1)+A'K^{2}+ \frac{1}{2}f_{11}\left(
5K^{3}-3KI(I+1)\right)
\nonumber \\
&+&f_{\mbox{\scriptsize qoc}}\frac{1}{I^{2}} \left(
15K^{5}-14K^{3}I(I+1)+ 3KI^{2}(I+1)^{2}\right) \ . \label{EKI}
\end{eqnarray}

We remark that a similar structure of the high order angular
momentum terms can be obtained by applying the method of contact
transformations \cite{BM75,Birss70} to a Hamiltonian including the
intrinsic particle motion and the Coriolis interaction \cite{Jolos05}.
The analysis in reference \cite{Jolos05} provides a deeper physical meaning
of the QORM Hamiltonian parameters in relation to the matrix elements
of the intrinsic angular momentum operators in the intrinsic single
particle states.

Equation (\ref{EKI}) provides the main part of the collective rotation
energy corresponding to the QORM Hamiltonian. The off-diagonal
terms in the octupole Hamiltonian (\ref{Hoct}) [see
equations.~(\ref{HF1F2})] have to be taken into account through a
numerical diagonalization procedure. Since the octupole shape is
fixed with respect to the space orientation angles, only three
off-diagonal terms, $\hat{H}_{F_{1}(2)}$, $\hat{H}_{F_{2}(1)}$ and
$\hat{H}_{F_{2}(2)}$ are considered to contribute to the structure
of the spectrum  \cite{octahed01}. Further, one may keep only
terms mixing states with neighboring $K$ values, i.e. $K\pm 1$,
omitting the term $\hat{H}_{F_{2}(1)}$ (with  $K\pm 2$ mixing)
whose contribution is essentially smaller. Then, noticing that the
matrix elements of the remaining terms $\hat{H}_{F_{1}(2)}$ and
$\hat{H}_{F_{2}(2)}$ have similar angular momentum dependences
(see the appendix in reference \cite{octahed01}) we realize that the
use of any one of them could provide a reasonable non-diagonal
octupole contribution to the energy of the system.

The model spectrum is obtained by minimizing the energy in the
diagonal part of  $\hat{H}_{\mbox{\scriptsize QORM}}$,
equation~(\ref{EKI}), with respect to $K$ and by diagonalizing the
total Hamiltonian (with the appropriate non-diagonal terms) in the
set of states $|IK\rangle$. The energy-minimum values of the
quantum number $K$, resulting from the minimization procedure,
gradually increase with the increase of $I$, forming a sequence of
the type $K=0,0,1,1,2,2,...$ for $I=1,2,3,4,5,6,...$. The increase
of $K$ after two consecutive equal values by a unit (a discrete
jump) generates an odd-even ($\Delta I=1$) staggering effect in
the collective band structure. The regularity of the $K$- sequence
depends on the behavior of the energy (\ref{EKI}) as a function of
$K$ and of the respective change of its minimum with $I$ (see figure
4). The numerical analysis of this behavior shows that after a
range of regularly ordered sequence, at some value of $I$ there is
an appearance of a single $K$- value $K=K_s$. For example one may
have $K=...,K_s-1,K_s-1,K_s,K_s+1,K_s+1,...$ The presence of such
an irregularity in the sequence causes an irregularity in the
$\Delta I=1$ staggering pattern. As a result the phase of the
oscillations is changed [the sign of the staggering quantity
(\ref{stag}) is inverted] and a ``beat'' staggering region is
formed. The angular momentum values where such irregularities
appear depend on the model parameters entering the energy
expression (\ref{EKI}). A detailed justification and specific
tests related to the appearing yrast $K$- sequences are given in
sec. VI of reference \cite{octahed01}.

In such a way the QORM Hamiltonian explains the appearance of
``beats'' in the staggering patterns of nuclear octupole bands. It
has been demonstrated that QORM is capable to reproduce the
``beat'' staggering effects in the high angular momentum regions
of alternating parity spectra in light actinide nuclei
\cite{Kyoto01,RILA02}. We should remark that the above presented
formalism can not describe alone the lowest levels of alternating
parity bands where the soft (oscillating) octupole mode dominates
the odd-even angular momentum shift.

\section{Description of the complete alternating parity band
structure}

The application of the formalism of the ``soft'' octupole mode
(given in section 2) and the model Hamiltonian of QORM
(section 3) allows us to describe the complete structure of nuclear
alternating parity bands. We take the collective model energy of
the quadrupole--octupole deformed system in the form
\begin{equation}
E_{\mbox{\scriptsize coll}}(I)=E_{\mbox{\scriptsize osc}}(I)
+E_{\mbox{\scriptsize QORM}}(I) \ , \label{Ecoll}
\end{equation}
where $E_{\mbox{\scriptsize osc}}(I)$, is given in
equation~(\ref{Elowvib}) and determined through the numerical solution
of the Schr\"{o}dinger equation for the potential (\ref{octfix})
as explained in section 2. It depends on the constant $E_0$, the
deformation $\beta_{\mbox{\scriptsize min}}$ in the potential
minimum, and the inertial parameters $d_1$ and $d_2$ of the
centrifugal term in equation~(\ref{Ubeta}). We take the same value for
the mass parameter $B_3=200$ $\hbar^2$/MeV in all nuclei under
study, as done in \cite{Jolos94}. The second term in
equation~(\ref{Ecoll}), $E_{\mbox{\scriptsize QORM}}(I)$, represents
the rotation energy of the quadrupole--octupole shape determined
by $\hat{H}_{\mbox{\scriptsize QORM}}$, equation~(\ref{Hgen}), within
the procedure explained in the end of section 3 and presented in more
details in \cite{octahed01}. Here we consider that
$E_{\mbox{\scriptsize QORM}}(I)$ depends on the quadrupole
parameters $A$ and $A'$, the octupole shape parameters $f_{11}$
and $f_{12}$ [see equation~(\ref{EKI}) and equations~(\ref{HF1F2})], and the
parameter of quadrupole--octupole interaction,
$f_{\mbox{\scriptsize qoc}}$, equation~(\ref{Hqoc}).

\begin{table}
\caption{Theoretical and experimental energy levels (in keV) of
the octupole bands in $^{224}$Ra, $^{226}$Ra, $^{224}$Th and
$^{226}$Th. The respective parameter values are given in figures
4-7. References for the experimental data are given in the
respective columns. The RMS deviations (in keV) are given at the
bottom.}
\bigskip
\begin{center}
\begin{tabular}{ccccccccc}
\hline\hline
\\
& \multicolumn{2}{c}{$^{224}$Ra} & \multicolumn{2}{c}{$^{226}$Ra}
& \multicolumn{2}{c}{$^{224}$Th} & \multicolumn{2}{c}{$^{226}$Th} \\
$I$&  th    &exp \cite{Cocks97}&   th   & exp \cite{Cocks97}
   &  th    &exp \cite{Th224}  &   th   & exp \cite{Th226}          \\
\hline
\\
 1 &  245.6 &  215.9 & 281.8 & 253.7& 241.4& 251.0& 215.2& 230.4    \\
 2 &   99.1 &   84.5 &  63.2 &  67.7& 110.2&  98.1& 105.2&  72.2    \\
 3 &  298.6 &  290.5 & 352.4 & 321.5& 327.0& 305.3& 312.6& 307.5    \\
 4 &  276.3 &  251.0 & 210.7 & 211.7& 294.4& 284.1& 237.4& 226.4    \\
 5 &  415.8 &  433.1 & 441.7 & 447.0& 466.5& 464.5& 454.6& 450.5    \\
 6 &  490.6 &  479.6 & 421.5 & 416.7& 540.3& 534.7& 445.1& 447.3    \\
 7 &  619.4 &  641.0 & 599.6 & 627.2& 692.9& 699.5& 654.8& 657.9    \\
 8 &  743.7 &  755.7 & 668.1 & 669.6& 829.2& 833.9& 715.3& 721.9    \\
 9 &  888.3 &  906.9 & 828.2 & 858.2& 991.2& 997.7& 914.2& 923.1    \\
10 & 1043.5 & 1069.4 & 947.5 & 960.3& 1163.7&1173.8& 1032.3& 1040.3 \\
11 & 1207.9 & 1221.7 &1108.9 &1133.5& 1343.9&1347.3& 1227.5& 1238.4 \\
12 & 1387.9 & 1414.7 &1262.7 &1281.6& 1541.2&1549.8& 1388.6& 1395.2 \\
13 & 1570.0 & 1578.3 &1434.5 &1448.0& 1739.2&1738.7& 1588.2& 1596.0 \\
14 & 1771.5 & 1788.5 &1612.7 &1628.9& 1955.2&1958.9& 1778.7& 1781.5 \\
15 & 1968.1 & 1969.4 &1796.2 &1796.5& 2167.8&2164.7& 1987.0& 1989.4 \\
16 & 2183.2 & 2188.7 &1993.8 &1998.7& 2397.9&2398.0& 2197.0& 2195.8 \\
17 & 2396.0 & 2388.8 &2187.4 &2174.9& 2621.1&2620.2& 2414.9& 2412.8 \\
18 & 2621.2 & 2613.1 &2400.0 &2389.8& 2861.0&2864.0& 2637.6& 2635.1 \\
19 & 2847.3 & 2831.7 &2602.1 &2579.3& &            & 2863.8& 2861.1 \\
20 & 3079.3 & 3060.2 &2821.8 &2801.1& &            & 3094.0& 3097.1 \\
21 & 3316.0 & 3294.5 &3034.7 &3006.7& & & &        \\
22 & 3551.3 & 3527.3 &3258.3 &3232.7& & & &        \\
23 & 3795.9 & 3774.3 &3479.6 &3454.9& & & &        \\
24 & 4031.0 & 4012.4 &3704.2 &3685.6& & & &        \\
25 & 4280.6 & 4271.1 &3931.2 &3921.9& & & &        \\
26 & 4512.1 & 4513.2 &4153.9 &4158.2& & & &        \\
27 & 4760.6 & 4782.7 &4384.0 &4405.9& & & &        \\
28 & 4988.6 & 5031.4 &4601.8 &4650.7& & & &        \\
RMS& \multicolumn{2}{c}{$19.0$} & \multicolumn{2}{c}{$20.4$}
& \multicolumn{2}{c}{$8.1$} & \multicolumn{2}{c}{$9.8$}
\\
\hline \hline
\end{tabular}
\end{center}
\end{table}

We have applied the above presented formalism to the alternating
parity bands of the nuclei $^{224,226}$Ra and $^{224,226}$Th. We
choose them as the best examples for nuclear spectra carrying both
octupole deformation characteristics and  pronounced collective
rotation structure. In addition the spectra in $^{224,226}$Ra are
characterized with well developed beat staggering patterns. After
adjusting the model parameters to the respective experimental
data, we have obtained a good agreement between the theoretical
and experimental levels with a standard root mean square (RMS)
deviation of 19 keV for $^{224}$Ra,  20.4 keV for $^{226}$Ra, 8.1 keV
for $^{224}$Th and 9.8 keV for $^{226}$Th. This is shown in table 1,
where the theoretical and experimental energy values are compared.
The respective $\Delta I=1$ staggering patterns have been also
reproduced successfully. The theoretical and experimental
staggering patterns are compared in figures 4-7. The theoretical
patterns are extended providing model prediction for the structure
of octupole bands above the experimentally observed angular
momenta.

\begin{figure}[t]
\epsfxsize=12.cm   %or \epsfysize= HEIGHT cm
\centerline{\epsfbox{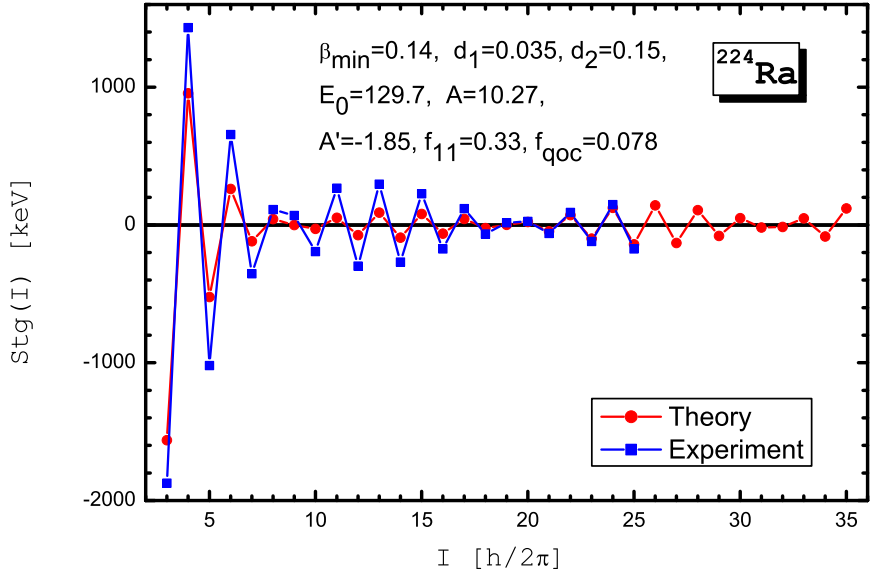}} \caption{$\Delta I =1$ staggering
patterns for the octupole band in $^{224}$Ra: experiment (data from
\cite{Cocks97}) and theory.  The parameters $B_3$, $d_1$ and $d_2$
are given in $\hbar^2$/MeV, MeV$^{-1}$ and MeV$^{-1}$, respectively,
$\beta_{\mbox{\scriptsize min}}$ is dimensionless, while the other
parameters are in keV.}
\label{fig:04}
\end{figure}

\begin{figure}[t]
\epsfxsize=12.cm   %or \epsfysize= HEIGHT cm
\centerline{\epsfbox{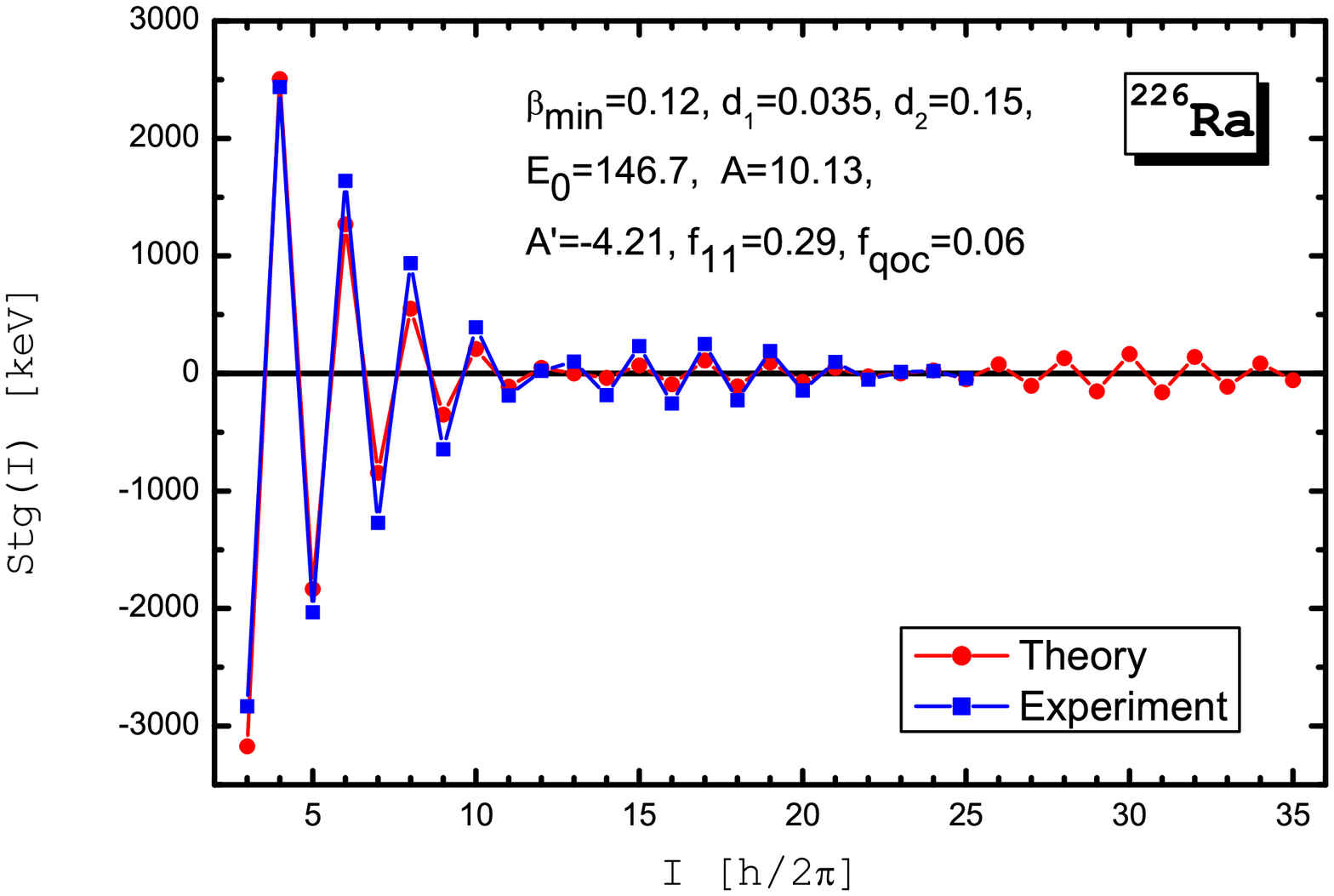}} \caption{The same as in figure 4,
but for the octupole band of $^{226}$Ra.}
\label{fig:05}
\end{figure}

\begin{figure}[t]
\epsfxsize=12.cm   %or \epsfysize= HEIGHT cm
\centerline{\epsfbox{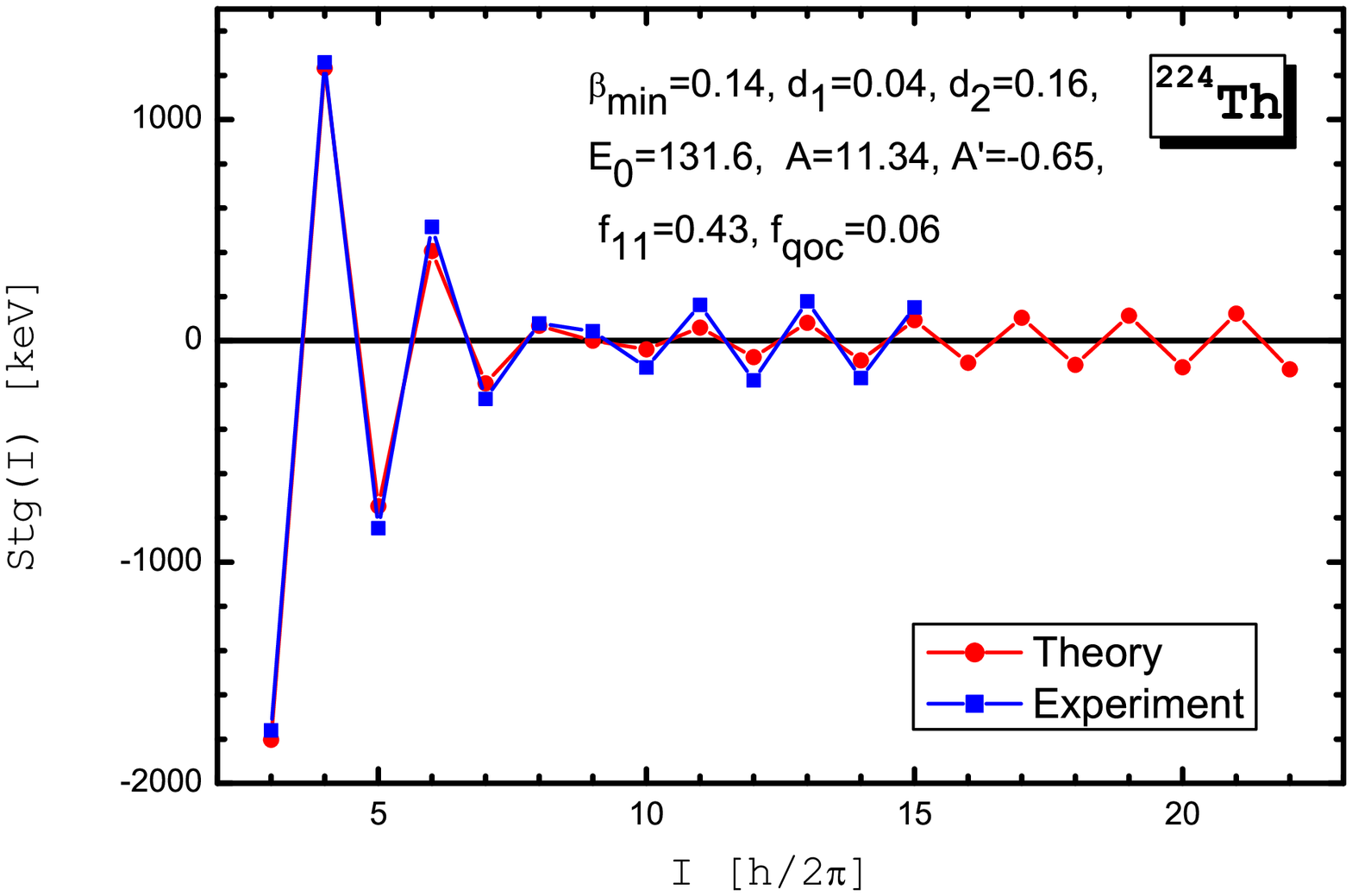}} \caption{The same as in figure 4,
but for the octupole band of $^{224}$Th. Data from \cite{Th224}.}
\label{fig:06}
\end{figure}

\begin{figure}[t]
\epsfxsize=12.cm   %or \epsfysize= HEIGHT cm
\centerline{\epsfbox{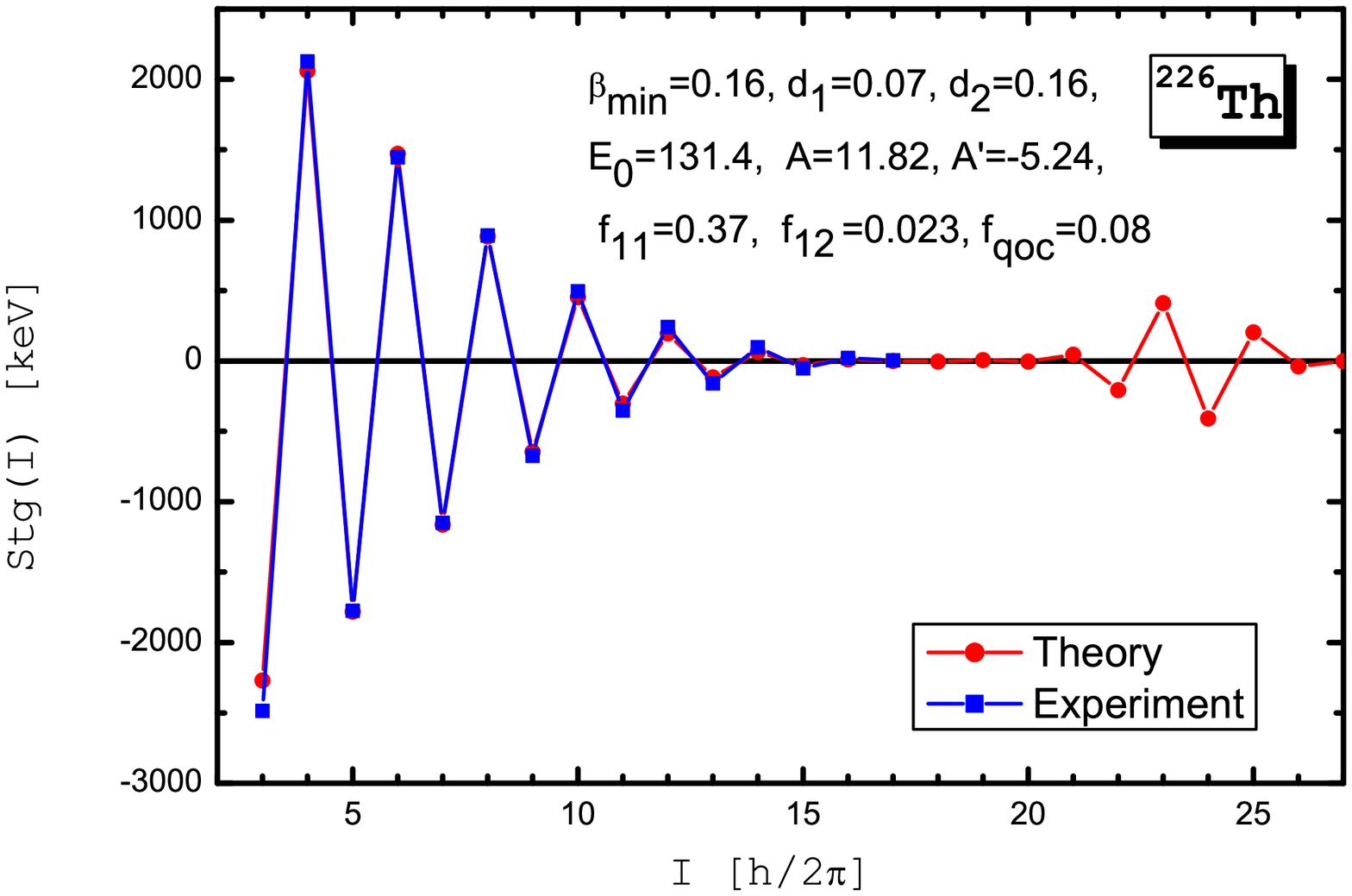}} \caption{The same as in figure 4,
but for the octupole band of $^{226}$Th. Data from \cite{Th226}.}
\label{fig:07}
\end{figure}

The following comments on the obtained results have to be made.

i) The model description reproduces correctly the points with near
zero staggering amplitude, which separate the different beat
regions in the staggering patterns. In this way the model
formalism clearly identifies the angular momentum region where the
two {\em separate sequences} of negative and positive parity
levels merge into a {\em single} octupole rotational band. We note
that both modes, the soft octupole and the stable
quadrupole--octupole, fit to each other quite consistently.

ii) As a consequence of i), our results show that the complicated
odd-even staggering structure of alternating parity bands can be
explained as the result of two consistently manifesting dynamical
effects in the nucleus. The one is the parity splitting effect due
to oscillations in the orientation (tunnelling effect) of the
octupole shape  and the other is the ''beat'' staggering effect
due to the angular momentum properties (with the specific
sequences of $K$-values described in the end of section 3) of the
rotating quadrupole--octupole shape. So, our model description
suggests that in the low angular momentum region of the spectrum
the staggering effect is mainly due to the parity effect, while in
the higher spin region the beat staggering structure of the band
is due to the rotating complex shape. In this aspect the first
``zero'' in the staggering amplitude can be attributed to a
transition angular momentum region where the nuclear collectivity
changes from the soft octupole to the stable rotation behavior.

iii) The successful reproduction of the second and the third beat
regions in the staggering patterns of $^{224}$Ra and $^{226}$Ra
provides a useful theoretical tool for the analysis and prediction
of collective band structure at high angular momenta. This is a
result of the model mechanism presented in the end of section 3.
According to the explanation given there, the presence of second,
third and further ``zeros'' in the staggering amplitude can be
interpreted as the result of respective irregularities in the
yrast $K$-sequences. Our calculations predict that
the third staggering region in $^{224}$Ra (figure 4) could be
completed near $I=30$ and a fourth region could be expected if
further experimental data are obtained. In $^{226}$Ra (figure 5),
again an additional third beat is expected near $I=30$. In
$^{224}$Th (figure 6) the prediction suggests that the second
staggering region could be continued in the region $I=15-20$,
while in $^{226}$Th (figure 7) the expectation is that a second
staggering region can take place for $I>20$. So, any further
experimental data in these spectra will be of great interest to
test the model predictions.

iv) We remark that the current calculations predict the presence
of a non-axial octupole deformation in $^{226}$Th with an
off-diagonal parameter value $f_{12}=0.023keV$. For the other
considered nuclei $f_{12}$ obtains zero values. The appearance of
a non-zero value of $f_{12}$ can be related to the recently suggested
shape phase transition between the regions of octupole deformations
and octupole vibrations \cite{AQOA}. It was shown that the nuclei
$^{226}$Ra and $^{226}$Th are closely placed to the transition point.
In this meaning the presence of triaxiality could be interpreted as
the sign of the transition from octupole deformation to octupole
vibration.

The following  comments on the applied model approach and its further
development should take a place here.

i) Although the total number of used model parameters looks large,
their numerical values in the different nuclei vary in quite
narrow limits (See the values shown in figures 4-7). We could keep
some of them ($d_1$ or $d_2$, $A$ and $A'$) as overall constants
similarly to the mass parameter $B_3$, without essential loss of
accuracy. However, at the current stage of study we prefer to
consider them as the output of the fitting procedure because of
their clear physical meaning. This consideration holds especially
for the parameter $\beta_{\mbox{\scriptsize min}}$ which provides
a direct model prediction for the octupole deformations of nuclei.
Actually it could be related to the QORM Hamiltonian parameter
$f_{11}$ through the nuclear moment of inertia. As a consequence a
possible change in $\beta_{\mbox{\scriptsize min}}$ with increasing
angular momentum could be allowed. Such a consideration would be a
step towards constructing a single model Hamiltonian where the same
degrees of freedom are incorporated and which possesses two different
limiting cases for low and high angular momenta.

ii) We note that further refinements of the considered approach
are possible. For example, it is known that the experimentally
observed parity shift in the light actinide nuclei decreases with
the angular momentum almost exponentially \cite{Jolos94}. We see in
figure 3(a) that in the first few levels with $I=1,2,3$ the theoretical
splitting does not decrease enough rapidly to fit this exponential
behavior, while for $I>3$ the fast decrease is already reached. We
notice that the behavior of the theoretical splitting for $I<3$ is
due to the change of the potential shape from a parabola at $I=0$ to
the double--well for $I>0$, which affects most strongly the energy shift
in $I=1$ and $I=2$. As a result relatively large discrepancies between
the theoretical and experimental energy values are observed at these
angular momenta (see table~1), with the related $R_4=E(L=4)/E(L=2)$
ratios (characteristics of collectivity) being also affected. This
problem can be fixed if the potential (\ref{Ubeta}) is modified so
as to obtain a double--well shape at $I=0$. The modification can be
achieved by applying in equation (\ref{Ubeta}) the term $d_0+X(I)$
instead of $X(I)$, where $d_0$ is a constant. A numerical analysis
shows that the behavior of the theoretical parity splitting and the
respective values of the low energy levels can be better reproduced
by means of the additional parameter $d_0$. So we suggest that such
efforts should be done in a further work after the electromagnetic
transition probabilities being considered (see below).

iii) The present formalism is capable of providing model
predictions on $B(E1)$, $B(E2)$ and $B(E3)$ reduced transition
probabilities between levels of alternating parity bands. Their
specific model behavior is related to the changing values of $K$
in the collective angular momentum states $|IK\rangle$. Our
preliminary analysis indicates that specific staggering effects in
the transition probabilities could be expected on this basis. On
the other hand the presence of $K$- mixing amplitudes coming from
the non-diagonal terms in the QORM Hamiltonian would restrict such
a staggering behavior. Therefore, the involvement of
electromagnetic transition probabilities in the study would make
the model more sensitive to the presence of non-axial octupole
deformations. In addition, the involvement of the ``intrinsic''
wave function determined in the solution of the Schr\"{o}dinger
equation is of importance in the model analysis of electromagnetic
transition probabilities. This study is the subject of further
work.

\section{Summary}
In conclusion, we propose a collective model formalism
incorporating the characteristics of the soft (oscillating)
octupole mode and the properties of the stable
quadrupole--octupole shapes in nuclei. This formalism is capable
to reproduce the entire structure of nuclear alternating parity
bands. We showed that it provides a successful description of the
octupole bands and their staggering structure in several light
actinide nuclei. The complicated ``beat'' staggering patterns are
explained as the result of the interplay between octupole shape
oscillations in a spin-dependent double-well potential and
rotations of the quadrupole--octupole shape. The model analysis
clearly indicates the transition between these two collective
modes. So, the implemented study suggests a specific dynamical
mechanism for the evolution of quadrupole--octupole collectivity.
The model predictions outline a possible way for development of
collectivity in the higher angular momenta with a presence of
further staggering regions. The considered dynamical mechanism can
govern a wide range of properties of nuclei with complex shapes.
\bigskip

\noindent {\Large \bf Acknowledgements}
\medskip

\noindent We thank Professor R V Jolos for valuable discussions and
advises. This work is supported by DFG and by the Bulgarian
Scientific Fund under contract No F-1502/05.

\end{document}